\documentclass[11pt]{JHEP3}
\usepackage{graphicx}
\usepackage{epsfig}
\usepackage{amssymb}
\input epsf.tex

\newcommand{\mueff}{\mu_{\rm eff}}
\newcommand{\beaa}{\begin{eqnarray*}}
\newcommand{\eeaa}{\end{eqnarray*}}
\newcommand{\bc}{\begin{center}}
\newcommand{\ec}{\end{center}}

\newcommand{\intx}{\int d^4 x\,}

\newcommand{\intvecx}{\int d^3 x\,}

\newcommand{\veck}{{\bf k}}

\newcommand{\vecx}{{\bf x}}       

\newcommand{\dl}{\delta}
\newcommand{\ep}{\epsilon}

\newcommand{\kp}{\kappa}
\newcommand{\lm}{\lambda}

\newcommand{\sg}{\sigma}

\newcommand{\ph}{\phi}

\newcommand{\Gm}{\Gamma}

\newcommand{\quart}{\frac{1}{4}}

\newcommand{\Tr}{\mbox{Tr}}

\newcommand{\phd}{\ph^{\dagger}}

\newcommand{\eela}[1]{\label{#1}\end{equation}}
\newcommand{\eeala}[1]{\label{#1}\end{eqnarray}}
\newcommand{\be}{\begin{equation}}
\newcommand{\ee}{\end{equation}}
\newcommand{\bea}{\begin{eqnarray}}
\newcommand{\eea}{\end{eqnarray}}
\newcommand{\dcp}{\delta_{\rm cp}}

\title{Simulations of Cold Electroweak Baryogenesis: 
Dependence on Higgs mass and strength of CP-violation }

\author{Anders Tranberg$^{a,b}$ and Jan Smit$^{c}$\\
$^{a}$) Department of Physics and Astronomy, University of Sussex,\\
Falmer, Brighton, East Sussex BN1 9QH, United Kingdom.\\
$^{b}$) DAMTP, University of Cambridge \\
Wilberforce Road, Cambridge, CB3 0WA, United Kingdom.\\
$^{c}$) Institute for Theoretical Physics, University of Amsterdam, \\
Valckenierstraat 65, 1018 XE Amsterdam, the Netherlands.\\
}


\preprint{ITFA-2006-17, DAMTP-2006-29}

\abstract{Cold electroweak baryogenesis was proposed as a scenario to bypass 
generic problems of electroweak baryogenesis within the 
Standard Model. In this scenario, baryogenesis takes place during an
electroweak symmetry breaking transition, 
which is also responsible for preheating after inflation. In the simplest 
modelling of the scenario, only 
two parameters remain undetermined: The Higgs mass
and the 
strength of CP violation. Using full real-time lattice simulations, we 
compute the dependence of the asymmetry on 
these parameters.
}

\begin{document}


\section{Introduction\label{introduction}}

Any scenario of baryogenesis aims at reproducing the observed matter-antimatter asymmetry of the Universe, usually quantified as the baryon-to-photon number density ratio \cite{Spergel:2006hy},
\be
\label{asymmetry}
\frac{n_{B}}{n_{\gamma}}= 6.1\times 10^{-10}.
\ee
In 
the original electroweak baryogenesis
scenario, this is achieved using 
Standard Model 
(SM)
physics (or electroweak physics of a supersymmetric extension of the SM), at a first order electroweak phase transition 
\cite{Kuzmin:1985mm} (see \cite{Rubakov:1996vz} for a review). Within the SM proper, the electroweak phase transition is a cross-over and cannot accommodate the required out-of-equilibrium conditions for successful baryogenesis 
\cite{Kajantie:1995kf}.
Furthermore,
(at least at finite temperature) the SM CP-violation in the CKM fermion mass matrix is insufficient by many orders of magnitude \cite{Shaposhnikov:1987tw,Shaposhnikov:1988pf}. 
In Cold Electroweak Baryogenesis, 
a period of
inflation is assumed to end at the electroweak scale
\cite{Garcia-Bellido:1999sv,Krauss:1999ng}, and 
subsequently
electroweak symmetry breaking 
takes place at zero temperature, 
is strongly out of equilibrium
through the process of tachyonic preheating, 
and 
is also responsible for the (re)heating of the Universe.
Further details on
different aspects of the scenario can be found in 
\cite{Copeland:2001qw,German:2001tz,vanTent:2004rc} 
(low-scale inflation), 
\cite{Skullerud:2003ki,Garcia-Bellido:2002aj,Diaz-Gil:2005qp} 
(electroweak tachyonic preheating), 
\cite{Smit:2002yg,Garcia-Bellido:2003wd,Tranberg:2003gi} 
(generation of the asymmetry),
\cite{Smit:2004kh} (SM CP-violation at zero temperature). 
This work is a continuation of \cite{Tranberg:2003gi}, where full lattice simulations of the electroweak transition including CP-violation were 
carried out.


\section{The SU(2)-Higgs model with CP-violation \label{su2higgs}}
We study the model described by the 
action:
\be
\label{lat_lagr}
S=
-\int d^{3}{\bf x}\,dt \bigg[\frac{1}{2g^{2}}\Tr F^{\mu\nu}F_{\mu\nu}+(D^{\mu}\phi)^{\dagger}D_{\mu}\phi
+ \ep 
+\mu_{\rm eff}^{2}(t)\phi^{\dagger}\phi+\lambda(\phi^{\dagger}\phi)^{2}+\kappa\,\phi^{\dagger}\phi\,
\Tr F^{\mu\nu}\tilde{F}_{\mu\nu}\bigg]\nonumber,
\ee
where $\ep$ is such that the energy density in the ground state is zero,
$\mueff$ is a time-dependent effective mass for the Higgs field and
$\kappa$ 
parametrises the strength of effective CP violation.
In Cold Electroweak Baryogenesis, Higgs symmetry breaking is triggered by a coupling to an 
inflaton.
In 
Inverted Hybrid Inflation \cite{Copeland:2001qw},
e.g.\
the one in \cite{vanTent:2004rc},
\be
\label{mueff}
\mu^{2}_{\rm eff}(t)\phi^{\dagger}\phi=\left[\mu^{2}-\lambda_{\sigma\phi}\sigma^{2}(t)\right]\phi^{\dagger}\phi,
\ee
where $\sg(t)$ is the time-dependent expectation value of the inflaton field.
As in \cite{Tranberg:2003gi} we will specialise to the instantaneous quench,
\be
\label{instantaneous}
\mu^{2}_{\rm eff}(t<0) = \mu^{2},\qquad \mu^{2}_{\rm eff}(t>0)=-\mu^{2},
\qquad
\ep = \mu^4/(4\lm).
\ee
The case of non-zero quench time will be treated in a separate publication \cite{CPtdep}.
A sufficiently rapid change of sign in $\mueff^2$ induces a spinodal instability with large occupation numbers, enabling us to use a classical approximation \cite{Garcia-Bellido:2002aj,Smit:2002yg}.  
 
The CP-violating term is to be thought of as
a lowest dimensional 
effective contribution from a theory beyond the SM 
\cite{Shaposhnikov:1987tw,Shaposhnikov:1988pf}. 
The parameter $\kp$ is dimensionful and can be written in terms of
a dimensionless parameter $\dcp$ as
\be
\label{kappa}
\kappa=\frac{3\delta_{\rm cp}}{16\pi^{2}m_{W}^{2}}.
\ee
For definiteness we have used $m_W$ as the mass scale in (\ref{kappa}),
$m_W = g v/2$, with $v$ the vacuum expectation of the Higgs field,
$v^2 = \mu^2/\lm$.
Experimentally, $m_{W}\simeq 80$ GeV and 
$v=246$ GeV,
which fixes the gauge coupling to be $g\simeq 0.65$. We use $g=2/3$ corresponding to $m_{W}=82$ GeV. 
We will allow the Higgs mass to vary relative to the W mass, determining the Higgs self-coupling through 
\be
\label{massratio}
\left(\frac{m_{H}}{m_{W}}\right)^{2}= \frac{8\lambda}{g^{2}}.
\ee
For the cases considered here, $m_{H}^{2}/m_{W}^{2}=2,3,4$. 

We choose $\dcp$ to be in the 
interval $\dcp=[0,1]$. The aim is to interpolate to very small values of $\dcp$ since, as we will see, to reproduce the observed 
baryon asymmetry
we will need $\dcp=\mathcal{O}(10^{-5})$. Ideally, we are looking for a linear regime at small $\dcp$. In \cite{Tranberg:2003gi}, we found that the dependence is non-linear for the range of $\dcp$ used there, and the present work zooms in on the interval between zero and the first non-zero value of $\dcp$ in \cite{Tranberg:2003gi}.


\subsection{Equations of motion and observables\label{eom}}
The action is discretised on the lattice and the classical equations of motion are derived (for details on the lattice implementation, see \cite{Tranberg:2003gi}). In the continuum, they read:
\bea
\label{EOM}
\left(D_{\mu}D^{\mu}+\mu^{2}_{\rm eff}(t)-2\lambda\phi^{\dagger}\phi-\kappa\Tr F^{\mu\nu}\tilde{F}_{\mu\nu}\right)\phi &=& 0,\\
D_{0}\left(\frac{1}{g^{2}}E^{a}_{k}-2\kappa\phi^{\dagger}\phi B^{a}_{k}\right)-\epsilon_{klm}D_{l}\left(\frac{1}{g^{2}}B^{a}_{m}+2\kappa\phi^{\dagger}\phi E^{a}_{m}\right)+j^{a}_{k}&=&0.
\eea
with $E_{k}^{a}=F^{a}_{k0}$, $B_{k}^{a}=\epsilon_{klm}F^{a}_{lm}/2$, $D_{l}$ is the adjoint covariant derivative $D^{ac}_{\mu}=\delta^{ac}\partial_{\mu}+\epsilon_{abc}A^{b}_{\mu}$ and $j_{\mu}^{a}$ is the $SU(2)$ current from the Higgs field,
\be
\label{higgscurrent}
j_{\mu}^{a}=i\left(D_{\mu}\phi\right)^{\dagger}\frac{\tau^{a}}{2}\phi-i\phi^{\dagger}\frac{\tau^{a}}{2}D_{\mu}\phi.
\ee
The Gauss constraint,
\be
\label{gaussconstraint}
D_{k}\left(\frac{1}{g^{2}}E^{a}_{k}-2\kappa\phi^{\dagger}\phi B^{a}_{k}\right)+j_{0}^{a}=0,
\ee
should be imposed on the initial condition, and will then be conserved by the equations of motion.

As was 
mentioned in \cite{Tranberg:2003gi}, a 
good lattice implementation of $F\tilde{F}$ results in implicit equations of motion, which require iterative solving. In combination with the 
number of terms arising from 
$F\tilde{F}$,
this means an increase of computer running time by roughly a factor
10. The work presented here amounts to about 30 CPU-years. 

We use periodic boundary conditions with spatial volume $L^3$, and
study the evolution of the Higgs expectation value,
\be
\label{phisq}
\tilde{\phi^{2}}
=
\frac{1}{L^3}\int d^{3}x\,\frac{\phi^{\dagger}\phi}{v^2/2} ,
\ee
the Chern-Simons number,
\be
\label{NCS}
N_{\rm cs}(t)-N_{\rm cs}(0)=\frac{1}{16\pi^2}\int dt\int d^{3}x\,\Tr F^{\mu\nu}\tilde{F}_{\mu\nu},
\ee
and the Higgs winding number, 
\be
\label{NW}
N_{\rm w}=
\frac{1}{24\pi^2}
\int d^{3}x
\, \ep_{ijk}
\Tr\bigg[(\partial_{i}U)U^{\dagger}(\partial_{j}U)U^{\dagger}(\partial_{k}U)U^{\dagger}\bigg],~~ U=\frac{\Phi}{\frac{1}{2}\Tr\Phi^{\dagger}\Phi},~~ 
\Phi=(i\tau_2\ph^*,\phi)\nonumber.
\ee
$N_{\rm w}$ is integer and can only change if there is a zero of the Higgs field. Such a zero is energetically unfavourable. Once the Higgs field 
has settled near the bottom of the potential 
($\tilde{\ph^2}\approx 1$)
and the temperature is 
relatively low ($\approx 50$ GeV \cite{Skullerud:2003ki}), 
no further changes should be seen in Higgs winding. 
In equilibrium, at 
sufficiently
high temperature, winding number changing transitions 
occur when the Higgs field goes through sphaleron-like configurations.
(Way) out of equilibrium and in the presence of a large number of Higgs zero's, winding number can change readily. Winding and unwinding during the tachyonic electroweak transition was studied in \cite{Garcia-Bellido:2003wd,vanderMeulen:2005sp}.

$N_{\rm cs}$ is integer in the gauge vacua, and equal to $N_{\rm w}$. At finite temperature (or finite energy density, out of equilibrium), it need not be integer and may be very different from $N_{\rm w}$. Still, as the system thermalises to a low temperature, we would expect the Chern-Simons number to relax to a value close to the winding number. This is what we see happening for late times. We will use this fact to simulate only until the Higgs winding has settled and the transition is over. This winding will then tell us what the asymmetry in Chern-Simons number would be, had we waited for it to settle.


\subsection{CP symmetric initial conditions\label{initial}}
We initialise our Higgs field using the 
``just the half''
method 
\cite{Smit:2002yg}
(also used in \cite{Salle:2001xv,Rajantie:2000nj,Copeland:2002ku,Skullerud:2003ki,Tranberg:2003gi}). 
An ensemble of configurations is generated reproducing the quantum
two-point functions in the vacuum before the 
quench.
\be
\label{initcond}
\langle\phi_{\bf k}\phi_{\bf k}^\dagger\rangle=\frac{1}{2\omega_{\bf k}},\quad\langle\pi_{\bf k}\pi_{\bf k}^\dagger\rangle=\frac{\omega_{\bf k}}{2},\quad \omega_{\bf k} = \sqrt{\mu^{2}+k^{2}},
\ee
with 
$\ph_\veck = L^{-3/2} \intvecx e^{-i\veck\vecx}\, \ph(\vecx)$,
and similarly for $\pi$.
Gauge fields $A_{i}$ are zero initially, with their canonical momenta $E_{i}$ determined through the Gauss constraint.
 
The ensemble of initial configurations is CP-symmetric. However, in a numerical simulation one only has a finite number of initial configurations available. Let $\overline{N} \equiv \frac{1}{M}\sum_{j=1}^M N_j$,
be the numerical estimate for $\langle N\rangle$,
where $N=N_{\rm cs}$ or $N_{\rm w}$ and $M$ is the number of initial configurations. Even for $\dl_{\rm cp} = 0$, $\overline{N}$ is typically non-zero because of statistical fluctuations. 
In a plot of $\overline{N}$ versus $\dl_{\rm cp}$ this leads to large uncertainties in the slope $d \overline{N}/d\dl_{\rm cp}$ near the origin.
Previously 
\cite{Tranberg:2003gi} we dealt with this problem by using the same series of pseudo random numbers for $\dl_{\rm cp} =0$ and $\dl_{\rm cp} \neq 0$. 
Here we avoid it by including the CP-conjugate configuration 
with every randomly generated initial configuration. 

We define the observables 
\be
\label{kis0}
\Delta_{N_{\rm cs}}=\frac{N_{\rm cs}+N_{\rm cs}^{CP}}{2},\qquad \Delta_{N_{\rm w}}=\frac{N_{\rm w}+N_{\rm w}^{CP}}{2},
\ee
where  
$N,N^{CP}$ correspond to the values for a CP-conjugate pair of 
initial configurations, as in (\ref{CP_equiv}). 
Taking ensemble averages, we obviously have 
\be
\langle \Delta_{N_{\rm cs}}\rangle= \langle N_{\rm cs}\rangle,\qquad \langle \Delta_{N_{\rm w}}\rangle= \langle N_{\rm w}\rangle.
\ee
These observables have the advantage that they cancel out some of the statistical noise. In particular, $\Delta_{\rm W}$ takes integer and half-integer values,
which reduces fluctuations. Standard errors calculated in terms of 
$\Delta$ are smaller than for 
$N$.
We have
\be
\sigma^2_{\Delta}=\left\langle\left(\frac{N+N_{CP}}{2}\right)^{2}\right\rangle - \left\langle\frac{N+N_{CP}}{2}\right\rangle^{2}=\frac{1}{2}
\left(\sigma_{N}^{2}+\langle N N_{CP}\rangle-\langle N\rangle\langle N_{CP}\rangle\right),
\ee
where we 
used
\be
\sigma_{N}^{2}=
\langle N^{2}\rangle - \langle N\rangle^{2}=\langle N_{CP}^{2}\rangle - \langle N_{CP}\rangle^{2}.
\ee
This gives an error estimate of 
\be
\label{errordelta}
error^{2}_{\Delta} = \frac{\frac{1}{2}(\sigma_{N}^{2}+\langle NN_{CP}\rangle-\langle N\rangle\langle N_{CP}\rangle)}{M/2-1},
\ee
where $M/2$ is the number of pairs of configurations. 
It may be compared with using simply the observable $N$ with $M$ 
random initial configurations, 
\be
\label{error_old}
error^{2}_{\rm N} = \frac{\sigma_{N}^{2}}{M-1}.
\ee
In our case, $N\simeq -N_{\rm CP}$, in which case the cross correlator in (\ref{errordelta}) is large and negative
($\langle N_{CP}\rangle = \langle N\rangle$), reducing the error. 
In the limit of no cross-correlation and $M\gg 1$, the two error estimates (\ref{errordelta}), (\ref{error_old}) coincide.

Let a prime denote the operation of CP conjugation,
\be
\ph'(\vecx,t) = \ph'(-\vecx,t)^*,
\qquad
A'_k(\vecx,t) = -A_k(-\vecx,t)^T,
\ee
where $*$ denotes complex conjugation and $T$ denotes transposition. If
$\ph,A$ are a solution of the equations of motion with $\dl_{\rm cp}$, 
then $\ph',A'$ are a solution with $-\dl_{\rm cp}$ and CP-conjugate initial conditions. As before, let $N^{CP}(\dcp)$ denotes the final $N$ resulting from CP-conjugate initial conditions without changing $\dcp$. Since $N$ is odd under CP, 
it follows that
\be
\label{CP_equiv}
N^{CP}(\dcp)=-N(-\dcp),
\ee
as illustrated by the following diagram ($\dl\equiv\dcp$)
\be
\begin{array}{ccl}
\ph(\vecx,0)&\stackrel{t,\dl}{\longrightarrow}&N(\dl)\\
\downarrow\, CP&&\\
\ph'(\vecx,0)&\stackrel{t,-\dl}{\longrightarrow}&N'(-\dl)=-N(-\dl)
\end{array}
\ee
Expansion in $\dl$, 
\be
N = N_0 + N_1 \dl + N_2 \dl^2 + \mathcal{O}(\dl^3),
\qquad
\Delta = N_1 \dl + \mathcal{O}(\dl^3),
\ee
gives
\bea
\langle N\rangle &=& \langle N_1\rangle \dl + \langle N_2\rangle \dl^2 + \cdots,
\\
\sg_N^2 &=& \langle N_0^2\rangle + 2 \langle N_0 N_1\rangle \dl + 
\langle 2 N_0 N_2 + N_1^2\rangle \dl^2 + \cdots,
\\
\sg_{\Delta}^2 &=& \langle N_1^2\rangle \dl^2 + \mathcal{O}(\dl^4).
\eea
Note that the zeroth and first order terms are absent in $\sg_{\Delta}^2$,
suggesting a strong reduction in statistical noise for small $\dcp$.

Because of (\ref{CP_equiv})
we just need to run with $\pm\dcp$, rather than the actual CP-conjugate configurations. We 
checked this 
numerically.
In the following we will no longer distinguish between the exact
$\langle N\rangle$ and the numerical estimate $\overline{N}$.


\section{Numerical results}


%
\FIGURE{
\epsfig{file=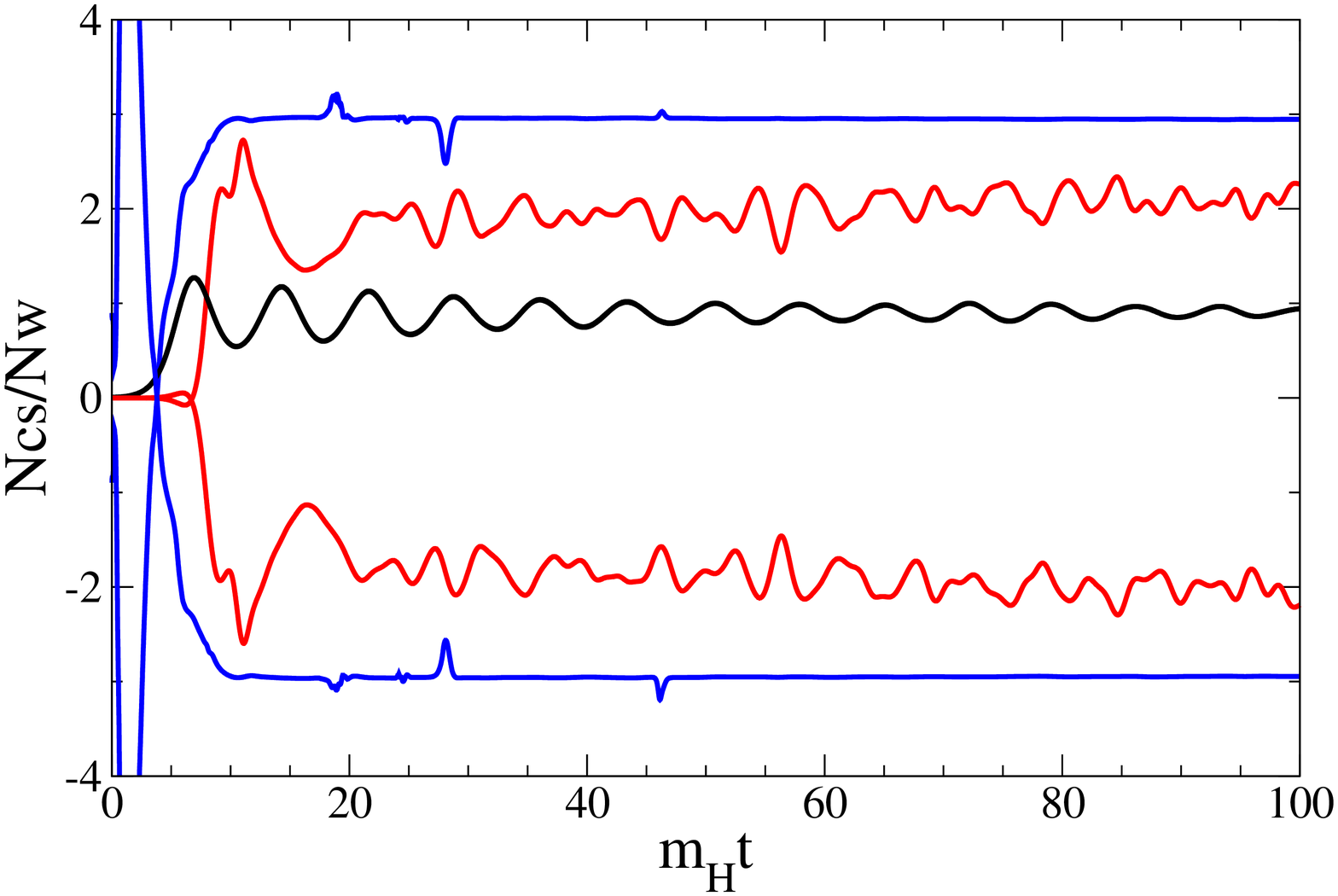,width=7cm,clip}
\epsfig{file=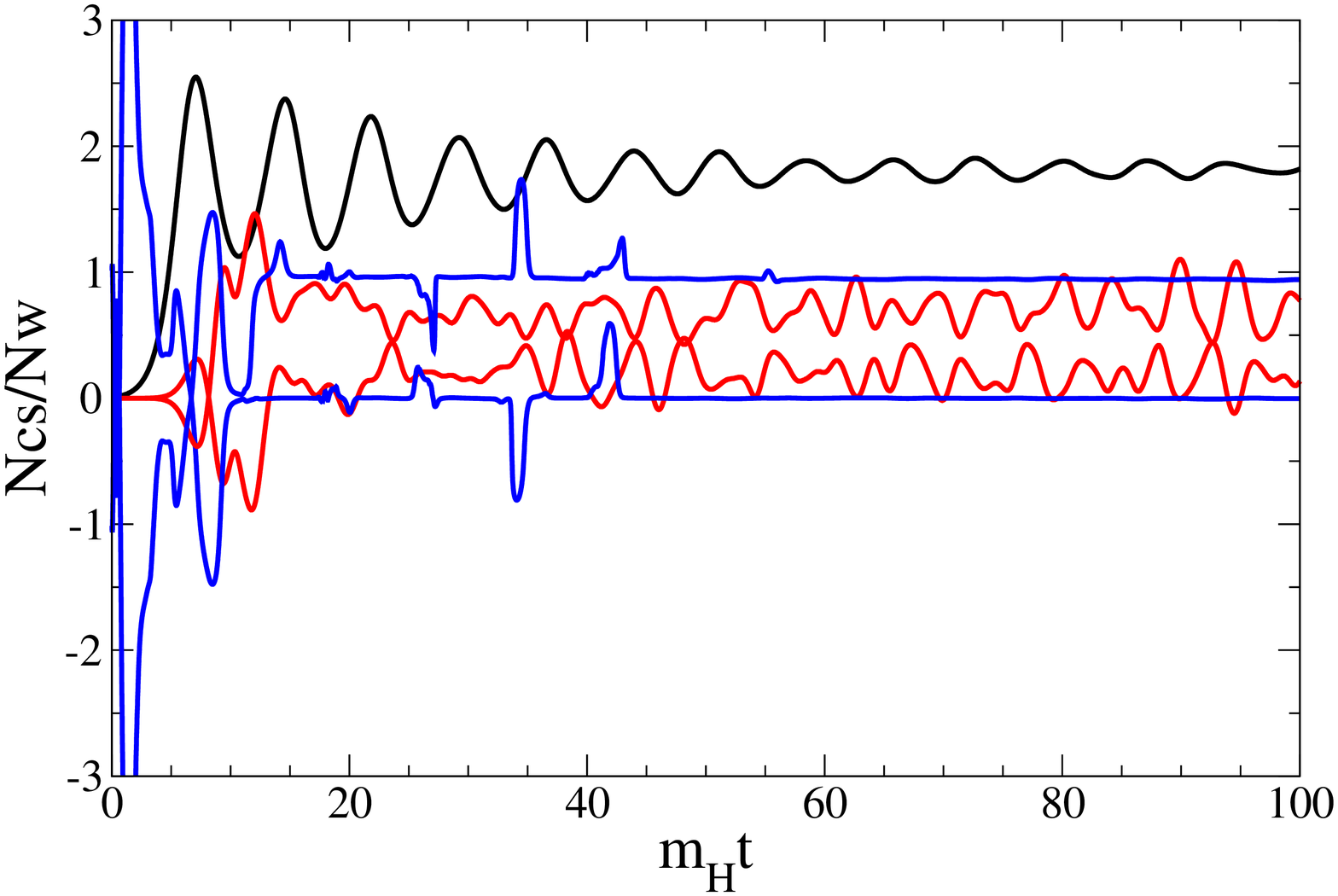,width=7cm,clip}
\caption{Example of CP conjugate pairs of configurations. Shown is the time evolution of $\tilde{\phi^{2}}$ (black), $N_{\rm cs}$ (red) and $N_{\rm w}$ (blue) for $\dcp=1$, $m_{H}/m_{W}=2$. In one case, the CP violation has very little effect (left) and $\Delta_{N_{\rm w}}=0$, in the other (right) the result is a net difference of $\Delta_{N_{\rm w}}=1$.}
\label{singletraj1}
}
In figure \ref{singletraj1}(left) we show the evolution from a single initial configuration, evolved with $\dcp=\pm 1$ and $m_{H}=2m_{W}$. For the two trajectories the Higgs field $\tilde{\phi}^{2}$ performs symmetry breaking in an identical way, settling near 1. At the same time, Chern-Simons number grows in an 
almost symmetric way. Shown here is $N_{\rm cs}(\dcp=1)$ and $-N_{\rm cs}(\dcp=-1)=N^{\rm CP}_{\rm cs}(\dcp=1)$. 
The Higgs winding number is truly symmetric after settling near time 10
(the glitches of magnitude less than one are discretisation errors). 
Obviously, this is a configuration pair with no generated asymmetry, $\Delta_{N_{\rm w}}=0$. Figure \ref{singletraj1}(right) is a similar pair of trajectories, but now $\Delta_{N_{\rm w}}=1/2$.

\FIGURE{
\epsfig{file=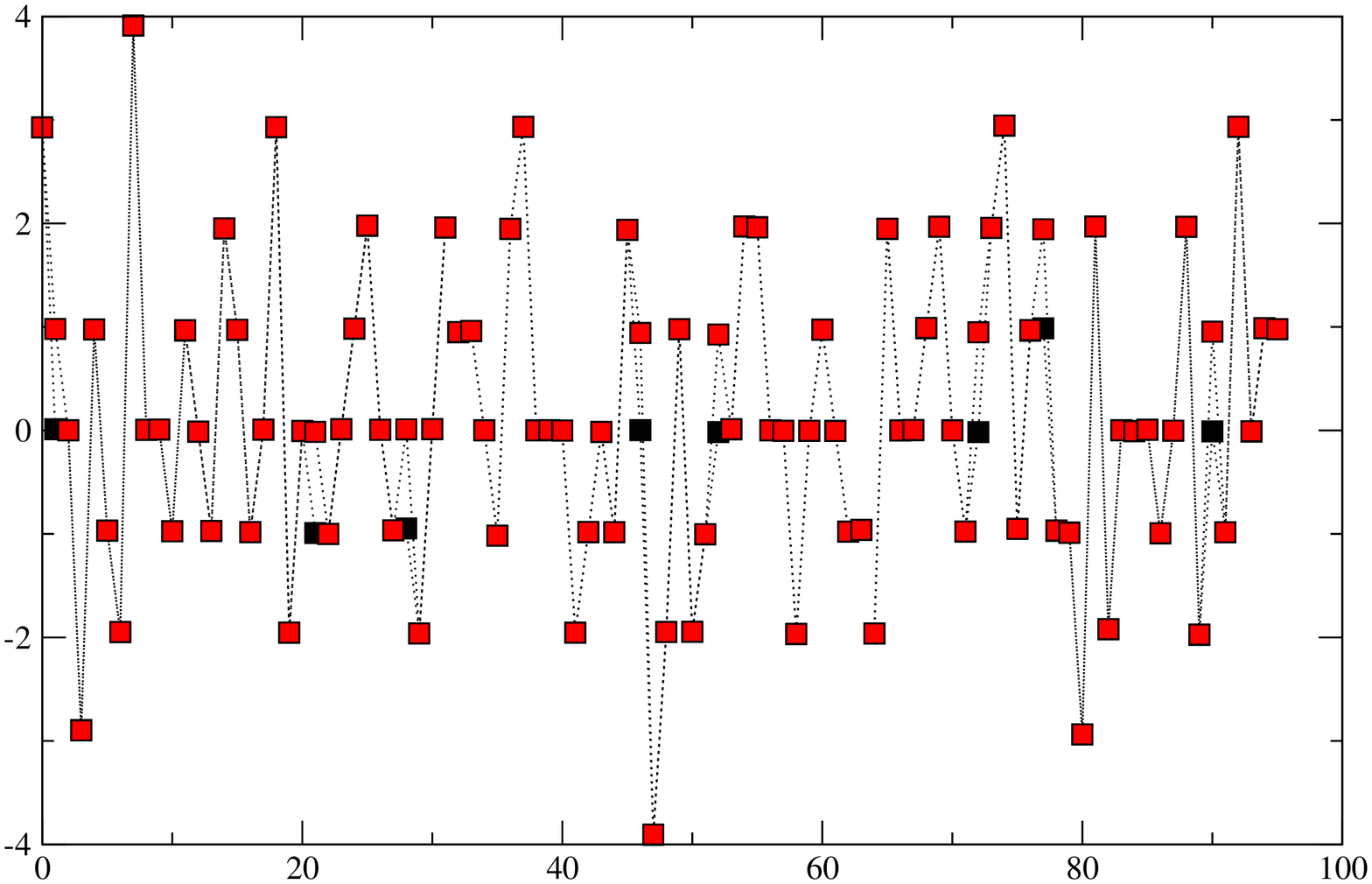,width=10cm,clip}
\caption{The final values of $N_{\rm w}$ for 
96 pairs of configurations. Black is $N_{\rm w}(\dcp)$, red $N_{\rm w}(-\dcp)=-N_{\rm w}^{\rm CP}(\dcp)$. When they are not on top of each other, a net asymmetry has been produced; $m_{H}/m_{W}=2$, $\dcp=1$.}
\label{distplot12_s4}
}
We run until $m_{H}t=100$.
Figure \ref{distplot12_s4} shows the resulting values of $N_{\rm w}$ for $+\dcp$ (black) and $-\dcp$ (red),
using an ensemble of 96 pairs of configurations. $\langle\Delta_{N_{\rm w}}\rangle$ is (a half times) the black minus the red values, averaged over 
the 96 pairs.

From 
now
on all results are for the ensemble averaged quantities $\langle N_{\rm w}\rangle$, $\langle N_{\rm cs}\rangle$.
Although the final asymmetry is what we are ultimately interested in, the full time evolution shows complicated features.
There is a linear regime during the first rolling off of the Higgs field. 
Then a non-linear back-reaction regime, where 
the behaviour of $\langle N_{\rm cs}\rangle$ can be described approximately in terms of diffusion under a time-dependent chemical potential \cite{Garcia-Bellido:1999sv,Garcia-Bellido:2003wd}.
A nice aspect of this description is that the dependence on the CP-violation is clearly linear. 
This regime ends when $\langle N_{\rm w}\rangle$ begins moving away from zero, to settle near the final value.
The change of $\langle N_{\rm w}\rangle$ is determined by the generated $\langle N_{\rm cs}\rangle$ as well as the availability of zeros of the Higgs field, and is as such a very complicated process.

In the following we will express time in units of the Higgs mass, 
$\tau=m_{H}t$.


\subsection{Initial rise
\label{initialrise}}
\FIGURE{
\epsfig{file=./pictures/bumpt0k12s4_v2.eps,width=8cm,clip}
\caption{Comparing the initial bump from the full simulation (red) to the estimate from the linearised and homogeneous equations of motion, eq. (\ref{linhom}) (green).}
\label{bump}
}
In the initial tachyonic instability of the Higgs field, low momentum modes dominate.
In \cite{Tranberg:2003gi} we solved for the early time evolution in 
the 
linear regime, making a homogeneity approximation and 
treating the CP violation as a perturbation. The result was that the generated asymmetry during the first roll-off of the Higgs field is given by:
\be
\langle N_{\rm cs}\rangle=
\frac{\sqrt{2}\dcp (Lm_{H})^{3}}{64\pi^4(1+c)^2}
\,
\frac{\langle\bar{B^{2}}\rangle}{m_{H}^{4}}
\,\tilde{\ph^2}.
\label{linhom}
\ee
The parameter $c$ 
is to be extracted from the growth of the magnetic field $B^{2}\propto \exp(2\sqrt{2}c\,\tau)$, and $B^{2}$ and 
$\tilde{\ph^2}$ are 
taken from the simulations. For the case 
$m_{H}/m_{W}=2$, $c\simeq 0.62$, for $m_{H}/m_{W}=\sqrt{2}$, $c\simeq
0.67$. Figure \ref{bump} shows the result for $\langle N_{\rm cs}\rangle$ from eq. (\ref{linhom}) (green) and the full simulation (red) for the case $m_{H}/m_{W}=2$. Notice that the scale is logarithmic. The discrepancy is 20 percent up to time 5.
When scaling the simulations for various $\dcp$ linearly with $\dcp$, the 
red curves fall on top of each other. 


\subsection{Early back-reaction: Asymmetric diffusion\label{asymdiff}}
The value of $\tilde{\ph^2}$ in figure \ref{singletraj1} indicates that
the system becomes non-linear after $\tau \simeq 5$. We may get some insight into the 
early
subsequent behaviour by considering the diffusion of Chern-Simons number under the influence of a chemical potential \cite{Khlebnikov:1988sr,Garcia-Bellido:1999sv,Garcia-Bellido:2003wd}. 
Making an approximation in which $\phd\ph$ in the CP-violating term in the action (\ref{lat_lagr}) is replaced by its spatial average
and making a partial integration, exhibits a time-dependent chemical
potential-like interaction for Chern-Simons number
\be
\label{mu_ch1}
-\intx \kappa\,\phi^{\dagger}\phi\,\Tr F^{\mu\nu}\tilde{F}_{\mu\nu}
\to
-\int dt\, \kp \frac{v^2}{2}\, \tilde{\ph^2} \frac{d}{dt} \, 16\pi^2 N_{\rm cs}=
\int dt\, \mu_{\rm ch}^{\rm ncs} N_{\rm cs},
\ee
with
\be
\label{much}
\mu_{\rm ch}^{\rm ncs}(t) = \frac{6\,\dcp}{g^{2}}\frac{d}{dt}\,
\tilde{\phi^{2}}(t),
\label{chempot}
\ee
where we also used the definition (\ref{kappa}) of $\dcp$ and $m_W^2 = \quart g^2 v^2$.
The effective diffusion rate of Chern-Simons number is\footnote{In equilibrium, $\Gamma$ is
called the sphaleron rate, 
which describes the widening of 
the distribution of Chern-Simons number. In the present out-of-equilibrium context, 
it does not have this 
straightforward physical interpretation. 
}
\be
\label{diffusion_rate}
\Gamma= \frac{d}{dt}\,
\left(\langle N_{\rm cs}^{2}\rangle-\langle N_{\rm cs}\rangle^{2}\right).
\ee
The generated Chern-Simons number asymmetry 
is then deduced to be
\cite{Garcia-Bellido:1999sv,Garcia-Bellido:2003wd}
\be
\label{linres_ncs1}
\langle N_{\rm cs}\rangle(t)=\int_{0}^{t}dt'\,
\frac{\mu_{\rm ch}^{\rm ncs}(t')\Gamma(t')}{T_{\rm eff}},
\ee
where $T_{\rm eff}$ can be thought of as some effective temperature of the relevant low momentum modes. 

Because the gauge fields become large as the transition proceeds, the effective diffusion rate grows in time. Performing the integration directly from the time-dependent, numerically determined 
$\Gamma(t)$ and $\tilde{\phi^{2}}(t)$ 
(figure \ref{suscep}) and eq.\
 (\ref{linres_ncs1}), 
one reproduces not only the initial rise, but also the subsequent dip, resulting in an asymmetry with the opposite sign from 
the initial rise (figure \ref{linres}). Indeed, because the diffusion rate is larger towards the end of the transition 
($\tau\approx 7$), it conspires with the Higgs field oscillation (with negative slope, effective chemical potential) to qualitatively change the final outcome.
\FIGURE{
\epsfig{file=./pictures/gamma_phi_t0.eps,width=10cm,clip}
\caption{The 
effective diffusion rate 
$\Gamma(t)$, eq.\ 
(\ref{diffusion_rate}) (red), and the time derivative of the Higgs
field, proportional to   
$\mu_{\rm ch}^{\rm ncs}(t)$, 
eq.\ (\ref{chempot}) (black),
both in units of $m_H$.
}
\label{suscep}
}
\FIGURE{
\epsfig{file=./pictures/linrest0.eps,width=10cm,clip}
\caption{Comparing 
eq.\ (\ref{linres_ncs1}) (black) for $\langle N_{\rm cs}\rangle$ to the full simulation (red), 
$m_{H}/m_{W}=2$,
$\dcp = 1/8$, 1/4, 1/2, 3/4, 1. 
The red lines are
curves for all $\dcp$, simply rescaled to $\dcp=1$. 
The blue lines are $\langle N_{\rm w}\rangle$, also rescaled. 
Notice that the average winding number does not move until around
$\tau =10$, the time of the first minimum of $\langle \tilde{\phi}^{2}\rangle(t)$.}
\label{linres}
}
In this argument, $T_{\rm eff}$ is an adjustable parameter, in figure \ref{linres} taken to be $\simeq 8 m_{H}$. This corresponds to $T_{\rm eff}=1.3\,\textrm{TeV}$, which is quite large. Once chosen, the semi-quantitative agreement for different $\dcp$ is convincing. The evolution of $\langle N_{\rm cs}\rangle$ is described by eq.\ 
(\ref{linres_ncs1}) 
until $\tau\simeq 8-10$.

The agreement 
ends around time $\tau=10$, which is also when the Higgs winding begins to grow. 
Apparently, the linear-response
treatment for the Chern-Simons number cannot account for the dynamics of winding and unwinding. For this, only the full non-linear simulations give a correct picture.


\subsection{Intermediate times: Higgs winding creation\label{intermediate}}
As we have seen, up to $\tau=10$ the statistical treatment of the Chern-Simons number is 
quite successful; There is a net $\langle N_{\rm cs}\rangle$, but still a 
tiny
$\langle N_{\rm w}\rangle$. Energetically, the two are strongly favoured to 
end up near each other at later times. 
This means that one has to adjust to the other\footnote{A similar situation has been studied in \cite{Turok:1990in,Turok:1991zg}, where it was seen that 
in single trajectories
the relative size of the winding and Chern-Simons number `blobs' is 
an indicator 
whether $N_{\rm w}$ adjusts to $N_{\rm cs}$ or vice versa.
}. 

\FIGURE{
\epsfig{file=./pictures/ratio2_v2.eps,width=8cm,clip}
\caption{The time history of $\langle\tilde{\phi^{2}}\rangle$ (black), $\langle N_{\rm cs}\rangle$ (red) and $\langle N_{\rm w}\rangle$ (green) for $m_{H}/m_{W}=2$,
$\dcp = 1$.}
\label{ratio2}
}

Higgs winding only changes when there is a zero of the Higgs field. 
The 
average Higgs field $\tilde{\ph^2}$ 
continues to oscillate some time after the transition (figure \ref{ratio2}). 
When it is low, the probability of zeros in $\ph$ itself is high.
The creation and evolution of 
(near)
zeros was studied in \cite{vanderMeulen:2005sp}, where it was seen that they indeed act as nuclei for winding number change as well as sphaleron-like transitions. It was also seen, that multiple ``generations'' of 
(near) zeros are generated, corresponding to subsequent minima of the Higgs oscillations. First generation nuclei are the most numerous, 
subsequent generations are less populated.

The existence of such zeros suggests why in the first Higgs oscillation, around $\tau=12$ the Higgs winding is able to adjust to the Chern-Simons number (figure \ref{ratio2}). For late times the winding number can no longer change, except through true sphaleron transition, for which the time scale at these temperatures is very long compared to the time scale of the simulation. We can estimate it through
$\Gm_{\rm sph}\propto e^{-E_{\rm sph}/T}$, with $E_{\rm sph}$ the sphaleron energy 
of order 10 TeV or 60 $m_H$. At time $\simeq 100\,m_{H}^{-1}$, a Bose-Einstein fit to the particle distribution functions gives $T/m_{H}\simeq 0.4$ \cite{Skullerud:2003ki}, suggesting that the sphaleron rate is indeed very small at these times.
It also suggests that $T_{\rm eff}$ as extracted from the asymmetric diffusion (section \ref{asymdiff}) should be interpreted with care. At longer times, the Chern-Simons number will settle 
close
to the winding number value\footnote{This we have checked for a few configurations, running to $\tau=500$.}.


\subsection{Dependence on Higgs mass\label{dephiggsmass}}
The end result turns out to be very sensitive to the Higgs to W mass ratio.
Here we 
present results for 
$m_{H}=\sqrt{2}\,m_{W}$ and $\sqrt{3}\,m_{W}$. In the former case (figure \ref{allratios}, left), the overall sign is opposite to what we saw in figure \ref{ratio2}. 
In the latter (figure \ref{allratios}, right), we are apparently in an intermediate case, where although there is still the initial linear regime, the dynamics conspires to give a final asymmetry consistent with zero.
The equation of motion of the Higgs field depends on the time derivative of the Chern-Simons number, and 
the frequencies and phases of these oscillation can conspire to give asymmetries of opposite signs. 
\FIGURE{
\epsfig{file=./pictures/ratios2_v2.eps,width=7cm,clip}
\epsfig{file=./pictures/ratios3_v2.eps,width=7cm,clip}
\caption{The time history of $\langle\tilde{\phi^{2}}\rangle$ (black), $\langle N_{\rm cs}\rangle$ (red) and $\langle N_{\rm w}\rangle$ (green) for different mass ratios; $m_{H}/m_{W}=\sqrt{2}$ (left) and $\sqrt{3}$ (right);
$\dcp= 1$.}
\label{allratios}
}

The large dependence on the Higgs mass is reminiscent of the situation for the analogous Abelian-Higgs system in 1+1 dimensions \cite{Smit:2002yg}. There, we were able to span a much larger range of masses, and the resulting curve looked quite complicated (figure \ref{massdep} (left)). For details about similarities and differences between the two studies, see \cite{Smit:2002yg}. 
\FIGURE{
\epsfig{file=./pictures/1d_massdep.eps,width=7cm,clip}
\epsfig{file=./pictures/3d_massdep_v2.eps,width=7cm,clip}
\caption{The Higgs mass dependence of the asymmetry in the analogous model in 1+1 (left) and 3+1 (right) dimensions; 
$\dcp = 1$,
$\kappa$ is the analogue of $\dcp$. Lefthand plot from
  \cite{Smit:2002yg}. In the righthand plot, blue points are the
  simulation presented here for $\dcp=1$, red are the results from the
  fits to the $\dcp$-dependence (see below) and black is the
  $m_{H}/m_{W}=1$ result from \cite{Tranberg:2003gi}.
}
\label{massdep}
}
At present we do not have the computational resources to perform an equally thorough study in 3+1 dimensions. But using the result for $m_{H}=m_{W}$ from \cite{Tranberg:2003gi} to guide us\footnote{An error in the 
application of
$\dcp$ 
in
\cite{Tranberg:2003gi} has been corrected.  
Furthermore,
results of \cite{Tranberg:2003gi}
should be multiplied by $g^{2}=4/9$,
the initial conditions 
for the case $m_H=m_W$ at small $\kp$
were different
(`thermal', resulting 
in somewhat
smaller final results),
and the quoted value 
refers to $\langle N_{\rm cs}\rangle$ rather than $\langle N_{\rm w}\rangle$.}, we can tentatively draw a plot of the mass dependence, figure \ref{massdep} (right).


\subsection{Dependence on CP-violation\label{depcp}}
\FIGURE{
\epsfig{file=./pictures/kdep_s4_v2.eps,width=7cm,clip}
\epsfig{file=./pictures/kdep_s2_v2.eps,width=7cm,clip}
\caption{Final $\langle N_{\rm w} \rangle$ as a function of $\dcp$. 
Left: $m_{H}/m_{W}=2$, right: $m_H/m_W=\sqrt{2}$. The full red line is a linear fit, the dotted lines represent $\pm 1\sigma$ in the fitted slope.
}
\label{allk_s4}
}
The nonlinear behaviour at intermediate times might also destroy the linear
dependence of the final asymmetry on $\dcp$. To study this we
vary $\delta_{\rm cp}$ using the 
same initial configurations for all $\dcp$. 
To get meaningful errors in the case $m_H/m_W=2$, we had to increase the number of CP-conjugate pairs to 192. Figure \ref{allk_s4} shows the final value of the average winding number vs.\ $\dcp$
up to $\dcp=1$. 
Within errors,
the dependence is consistent with linear.
The fits in figure 
\ref{allk_s4} and the one final value from figure \ref{allratios} (right) lead to 
an asymmetry 
\bea
\label{asymmetryk3rs2t0}
\langle N_{\rm w}\rangle &=& (0.075\pm 0.006)\dcp, 
\qquad
m_{H}=\sqrt{2}\,m_{W},
\nonumber\\
\langle N_{\rm w}\rangle &=& (0.005\pm 0.020)\dcp, 
\qquad
m_{H}=\sqrt{3}\,m_{W},
\nonumber\\
\langle N_{\rm w}\rangle &=& (-0.0359 \pm 0.0040) \dcp, 
\quad
m_{H}=2m_{W}.
\eea
%


\section{Conclusion\label{finalasym}}
Given the final ensemble average of the winding number, we can make an estimate for the generated baryon asymmetry. We use 
\be
\frac{n_{B}}{n_{\gamma}} = 7.04\frac{n_{B}}{s},\quad s = \frac{2\pi^2}{45}g_{*}T^{3},\quad\frac{\pi^{2}}{30}g_{*}T^{4}=
\ep
=\frac{m_{H}^{4}}{16\lambda},
\ee
The entropy $s$ is given in terms of the reheating temperature $T_{\rm reh}$ and $g_{*}$ the number of relativistic degrees of freedom and the reheating temperature deduced from the initial energy density in the Higgs potential, $T/m_{H}\simeq 0.45$. 
We also assume, as discussed earlier, that the 
late times $\langle N_{\rm cs}\rangle$ will be equal to $\langle N_{\rm w}\rangle$ at the end of our simulation. 
We have
\be
\frac{n_{B}}{n_{\gamma}} = 7.04\frac{3\langle N_{\rm w}\rangle}{(Lm_{H})^{3}}\left(\frac{45}{2\pi^{2}}\right)\left(\frac{15}{\pi^{2}g^{2}}\right)^{-3/4}g_{*}^{-1/4}\left(\frac{m_{H}}{m_{W}}\right)^{3/2}.
\ee
With $Lm_{H}=27$, $g=2/3$ and $g_{*}=86.25$, this gives
\be
\frac{n_{B}}{n_{\gamma}} =0.32\times 10^{-3}\times\langle N_{\rm w}\rangle \left(\frac{m_{H}}{m_{W}}\right)^{3/2}.
\label{asymconversion}
\ee
Finally, using the numerical results (\ref{asymmetryk3rs2t0}),
\bea
\label{allasymmetries}
\frac{n_{B}}{n_{\gamma}}&=& (0.40\pm0.03)\times 10^{-4}\times \dcp,~~~ m_{H}=\sqrt{2}m_{W},\nonumber\\
&=& (0.04\pm0.15)\times 10^{-4}\times \dcp,~~~ m_{H}=\sqrt{3}m_{W},\nonumber\\
&=&-(0.32\pm0.04)\times 10^{-4}\times \dcp,~~~ m_{H}=2m_{W},\nonumber
\eea
Compared to \cite{Tranberg:2003gi} the CP-symmetric initial conditions and the emphasis on the Higgs winding number allowed us to get a much clearer signal without much larger statistics. This was necessary in order to zoom in on the range of $\dcp$ where the dependence is linear. In particular, we were able to pin-point the time at which the asymmetry is generated to the first minimum of the Higgs field evolution. This is when the 
average winding number is able to change
and accommodate the initial asymmetry in the Chern-Simons number.
In the range 
$\dcp=[0,1]$, the asymmetry is linear in $\dcp$, allowing us to 
interpolate to the very small values relevant for the observed asymmetry. To reproduce the observations (\ref{asymmetry}), we need 
$\dcp\simeq 2\times10^{-5}$ ($m_{H}=2\, m_{W}$). 
Presumably, $\dcp$ should be somewhat larger than this, when taking into account the dynamics of the inflaton, fermions and additional gauge fields, which may in various ways affect the dynamics of the $SU(2)$-Higgs system.  In particular, the assumption of an instantaneous quench leads to 
quite wild behaviour. In very slow quenches, the system may never be 
sufficiently out of equilibrium, and the asymmetry should be correspondingly small. The dependence on the quench time will be presented in a separate publication \cite{CPtdep}.

The mass of the Higgs field in the Standard Model is expected to be smaller than 
$200$ GeV $\simeq2.5\, m_{W}$. 
We have probed the allowed region and found a dramatic dependence on 
$m_H$. Whether or not this effect survives at finite quench times is not yet known, and it is clear from the semi-analytic linear treatment in section \ref{asymdiff}, that the generic sign of the asymmetry is 
opposite to that of $\dcp$ for $m_{H}=2\, m_{W}$. Still, both for $m_{H}=\sqrt{2}\, m_{W}$ and for $m_{H}=m_{W}$ \cite{Tranberg:2003gi} the final result has the opposite sign,
i.e.\ the same sign as $\dcp$.


\subsection*{Acknowledgements}
We thank 
Margarita Garc{\'{\i}}a-P{\'e}rez, Antonio Gonz{\'a}lez-Arroyo, Andres D\'{\i}az-Gil, Juan Garc\'{\i}a-Bellido
and Meindert van der Meulen 
for useful remarks and enjoyable discussions. 
A.T.\ is supported by PPARC SPG {\it``Classical lattice field theory''}. Part of this work was conducted on the SGI Origin platform using COSMOS Consortium facilities, funded by HEFCE, PPARC and SGI as well as on the SARA PC-cluster LISA. This work received support from FOM/NWO.

\bibliography{anderslit2JS2}

\end{document}